# Insights into new mechanosensitive behaviors of G protein-coupled receptors

Aakanskha J Shetty[1], Alexei Sirbu[1,2], Paolo Annibale[1]*
[1] School of Physics and Astronomy, University of St Andrews, North Haugh, KY169SS, St Andrews, United Kingom
[2] ISAR Bioscience, Munich-Planegg
*pa53@st-andrews.ac.uk

**6142 words excluding references and figure captions**

## Abstract

G protein-coupled receptors (GPCRs) represent a diverse and vital family of membrane proteins that mediate intracellular signaling in response to extracellular stimuli, playing critical roles in physiology and disease. Traditionally recognized as chemical signal transducers, GPCRs have recently been implicated in mechanotransduction, the process of converting mechanical stimuli into cellular responses. This review explores the emerging role of GPCRs in sensing and responding to mechanical forces, with a particular focus on the cardiovascular system. Cardiovascular homeostasis is heavily influenced by mechanical forces such as shear stress, cyclic stretch, and pressure, which are central to both normal physiology and the pathogenesis of diseases like hypertension and atherosclerosis. GPCRs, including the angiotensin II type 1 receptor (AT1R) and the β2-adrenergic receptor (β2-AR), have demonstrated the ability to integrate mechanical and chemical signals, potentially through conformational changes and/or modulation of lipid interactions, leading to biased signaling. Recent studies highlight the dual activation mechanisms of GPCRs, with β2-AR now serving as a key example of how mechanical and ligand-dependent pathways contribute to cardiovascular regulation. This review synthesizes current knowledge of GPCR mechanosensitivity, emphasizing its implications for cardiovascular health and disease, and explores advancements in methodologies poised to further unravel the mechanistic intricacies of these receptors.

## Introduction

G protein-coupled receptors (GPCRs) represent one of the most extensive and diverse families of membrane proteins in the human body (Maudsley et al., 2005). These receptors play crucial roles in transmitting extracellular signals into intracellular responses, mediating a wide range of physiological processes (Lefkowitz, 2013). These are versatile, seven-transmembrane-domain proteins that regulate a diverse array of intracellular signaling cascades in response to hormones, neurotransmitters, ions, photons, odorants and other stimuli (Hilger et al., 2018). GPCRs are thus attractive drug targets as they play an important role in physiology and disease. Among the many physiological processes they regulate, their role in the cardiovascular system is particularly significant, as they modulate blood pressure, heart rate, and vascular tone. Recent advances in cellular biomechanics have highlighted an additional, less-explored aspect of GPCR regulation: their ability to sense and respond to mechanical stress and stimuli (Hardman et al., 2023), (Wilde et al., 2022), (Storch et al., 2012). This mechanotransduction process, where mechanical stimuli are converted into cellular signals, represents a novel dimension of GPCR function that extends beyond their classical role as chemical signal transducers. This review will focus on recent results covering GPCR response to external mechanical stimuli in a cardiovascular setting.

Mechanical forces play a pivotal role in maintaining cardiovascular homeostasis. The heart, blood vessels, and other components of the cardiovascular system are constantly exposed to mechanical stimuli, including shear stress, cyclic stretch, and pressure. These forces are critical for normal physiological function and contribute to the pathophysiology of various cardiovascular diseases, such as hypertension, atherosclerosis, and heart failure (Garoffolo and Pesce, 2019). For decades, it was assumed that mechanical stimuli primarily exert their effects through mechanosensitive ion channels (Friedrich et al., 2012), integrins (Friedrich et al., 2012) (Shyy and Chien, 2002), cell-cell adhesion molecules or cytoskeletal filaments (Ingber, 2006). GPCRs are thus a relatively recent addition to the list of mechanotransducers. The $G_{q/11}$-coupled angiotensin II type 1 receptor (AT1R) was the first GPCR to qualify as a mechanosensitive receptor (Zou et al., 2004), suggesting that these receptors not only respond to traditional ligands but are also activated or modulated by mechanical forces such as stretch or shear stress (Hardman et al., 2023). This dual activation mechanism allows GPCRs to integrate mechanical and chemical signals, contributing to their complex role in cardiovascular physiology.

The mechanistic underpinnings of how mechanical forces activate non-adhesion GPCRs are still being elucidated, but several models have been proposed. One such model suggests that mechanical stress directly induces conformational changes in the receptor structure, particularly within the transmembrane helices, leading to receptor activation. This mechanical deformation could alter the receptor's interaction with intracellular G proteins and other signaling molecules, triggering downstream signaling cascades similar to those initiated by ligand binding (Wilde et al., 2022). Another model proposes that mechanical stimuli might modulate the interactions between GPCRs and other membrane molecules, such as lipids, which could in turn influence receptor function (Poudel and Vanegas, 2024, Candelario and Chachisvilis, 2012). Furthermore, emerging evidence indicates that mechanical activation of GPCRs can promote biased signaling, in favour of specific intracellular pathways over others (Tang et al., 2014) suggesting that mechanical activation of GPCRs could occur both in a ligand-dependent or -independent way, defining two distinct modes of mechanoactivation. This form of allostery could have important implications for cardiovascular function and disease, as it may lead to distinct physiological outcomes while receptors are undergoing canonical ligand-induced activation (Cullum et al., 2024, Sirbu et al., 2024a). The β2-AR plays a key role in regulating cardiovascular function by mediating the effects of the sympathetic nervous system on heart rate, contractility, and vascular tone. When activated by catecholamines like adrenaline and noradrenaline, β2-AR primarily causes vasodilation in the blood vessels, particularly in the skeletal muscles, heart, and lungs. This vasodilation reduces peripheral resistance, which helps decrease blood pressure and increases blood flow to these critical tissues during times of stress or exercise. In the heart, β2-AR activation contributes to a mild increase in heart rate and myocardial contractility, although this effect is less pronounced than that for the highly homologous β1-AR. The β2-AR's function in modulating vascular tone helps balancing systemic vascular resistance and supports the body's ability to respond to physical or emotional stress. Interestingly, two independent reports from this year have addressed these two modes of activation for the same receptor, namely the β2-adrenergic receptor (β2-AR) (Cullum et al., 2024, Sirbu et al., 2024a).

In this review, we shall briefly highlight the key findings related to GPCR mechanosensitive behavior, already summarized in several excellent recent reviews (Wilde et al., 2022) (Hardman et al., 2023) (Xiao et al., 2023), with a specific focus on the cardiovascular system, before addressing specifically the case of the β2-AR and some of the upcoming methodological developments that may allow the field to move further.

**Mechanism of GPCR signalling**

For most GPCRs, binding of the endogenous hormone or ligand leads to conformational changes at the cytoplasmic ends of the transmembrane (TM) segments that provide an interaction interface for cytosolic proteins including heterotrimeric G proteins, G protein-coupled receptor kinases (GRKs) and arrestins (Manglik and Kobilka, 2014). Structural comparative studies of class A GPCRs highlight common activation mechanism, underpinned by tilt, rotation, elongation of the transmembrane domains (TM) that create a network of interactions between specific residues that stabilize the receptors in each conformation (Hauser et al., 2021). These studies highlighted conserved motifs of receptor conformational change upon binding of an agonist to the orthosteric pocket, the most obvious being movements of TM3 (Classes B and C) and TM6 (A, B and C). According to the cubic ternary model, inactive [R] and active [R*] states of a receptor are in equilibrium. In absence of a ligand some critical amino acid residues in the TM domains interact with one another, keeping receptors preferentially locked in an inactive state, corresponding to conformations that typically do not favor G protein binding and are incapable of catalyzing G protein nucleotide exchange (Manglik and Kruse, 2017). Rather, the activated [R*] state represents conformations that enable the receptor to interact with heterotrimeric G proteins or other effectors. The binding of the agonist changes the helix-helix interactions in the GPCR pushing the receptor towards [R*] (Gether, 2000, Yasuda et al., 2008) (Zhang et al., 2009). To create a cavity on the receptor intracellular face that can accommodate the G protein, GPCR activation involves a rotation and displacement of transmembrane (TM) helix 6. For instance, in the case of the β2-AR, a structural prototype for a large fraction of class A GPCRs, $Cys^{265}$ which is located at the end of the third intracellular loop (IC3), at the cytoplasmic end of the transmembrane 6 (TM6) α-helix, is important for G protein coupling and the constitutive receptor activation (Ghanouni et al., 2001). TM5 also rotates away from the receptor, further enlarging the G protein binding cavity. This enables movement of helixes 5, 6, and 7, translating into conformational changes in the third cytoplasmic loop that subsequently activate G proteins (Mahoney and Sunahara, 2016). Agonist-induced activation of GPCRs triggers downstream effector signaling that mediates canonical pathways, such as G protein and β-arrestin signaling. The GRK - β-arrestin system is a dual-function system that not only mediates receptor desensitization but also activates independent signaling pathways(Jiang et al., 2022).

Approximately 34% of the drugs approved by the US Food and Drug Administration (FDA) target GPCRs and are used to treat cardiovascular, neuropsychiatric, neurodegenerative, metabolic, and inflammatory diseases(Hauser et al., 2017). Most of these drugs are small molecules that bind to the orthosteric ligand binding sites of the GPCR, altering the receptor conformation and modulating the intracellular signaling responses (Persechino et al., 2022). Allosteric modulators, unlike orthosteric drugs, bind to receptor sites that are evolutionarily less conserved and spatially distinct from orthosteric sites. They mediate responses by selecting or stabilizing specific conformations of GPCRs when an orthosteric ligand is bound. GPCRs are well-known examples of allosteric proteins because the ligand binding at the extracellular orthosteric site promotes the binding of the G protein at the cytoplasmic side of the GPCR. This coupled equilibrium of agonist and G protein increases the affinity of the ligand binding (Weis and Kobilka, 2018) (Pani et al., 2021). Small molecules can also serve as allosteric modulators, such as a positive allosteric modulator (PAM) cmpd-6, (Pani et al., 2021) that enhances the pharmacological activity of carvedilol, at β2-AR and β1-AR sites. Carvedilol is a drug used in the treatment of cardiac dysfunction, and the modulation by cmpd-6 to stimulate β2-AR mediated extracellular signal–regulated kinase (ERK) phosphorylation in a β-arrestin dependent manner stabilizes the carvedilol-bound distinct conformation of β2-AR and inhibits β2-agonist coupled Gαs-protein stimulation and cAMP signaling. This allosteric modulation results in cardioprotective effects in heart failure. Molecules from the membrane environment also serve as typical allosteric modulators of receptor function(Zocher et al., 2012). The receptor can thus be influenced either by the general physicochemical properties of the membrane bilayer and interactions with specific lipids. The first group includes factors such as lateral pressure of the membrane bilayer, local membrane curvature, and the hydrophobic mismatch between GPCRs transmembrane domains and the lipid bilayer(Baccouch et al., 2022, Jones et al., 2020). All these factors can influence the receptor's energetic landscape and potentially shift its conformation. While the direct effect of these factors on GPCR conformation remains understudied, they are strongly dependent on the lipid composition of the membrane, which can be studied in controlled environments, such as lipid nanodiscs(Gunsel and Hagn, 2022). For instance, studies on ligand binding and structural dynamics have shown that specific phospholipids can modulate ligand binding affinity and receptor activity for the β2-AR(Dawaliby et al., 2016) and the ghrelin receptor(Damian et al., 2021). A key player in regulating both membrane properties and receptor function is cholesterol. As a vital component of all eukaryotic membranes, it plays a central role in the regulation of membrane fluidity and lipid organisation(Rog et al., 2009). Furthermore, cholesterol appears in most of the currently available X-ray and cryo-EM GPCR structures, residing in various locations(Taghon et al., 2021). Although a comprehensive understanding of its role in GPCR function has yet to be achieved, in the case of the β2-AR cholesterol has been shown to limit conformational variability and increase mechanical stability of the receptor by single-molecule force spectroscopy(Zocher et al., 2012) and atomistic simulations(Manna et al., 2016).

**Mechanical stress-induced activation of Class A GPCR signaling: the prototypical case of the AT1R**

In the light of such mechanisms of receptors activation, it is reasonable to argue that either direct or indirect mechanical stimuli acting on one or more domains of an individual receptor may either alter its conformation or modulate how it responds to a canonical (e.g. ligand-mediated) activation. Mechanical stimuli may not induce the same conformational changes in the receptor structure that orthosteric ligands cause. However, it has been established that they can induce different active states or allosterically modulate the canonical signaling (Hardman et al., 2023). While the sources of mechanical stimuli acting on cells in a physiological setting are extremely broad, the result at the receptor level would be a change in the forces exerted on individual receptor domains.

One of the prototypical cases where those effects have been initially studied is the angiotensin type 1 receptor (AT1R). The AT1R couples to numerous heterotrimeric G proteins (primarily $G_{q/11}$ but also $G_i$, $G_{12}$ and $G_{13}$). However, it can also prominently signal via β-arrestins. In fact, the AT1R was the first receptor for which a biased ligand was described over 20 years ago(Holloway et al., 2002).

AT1R plays an important role in mediating the external load into intracellular responses. When cardiomyocytes are stretched in vitro, they release Ang II (AII) which induces hypertrophic responses, indicating an indirect form of signaling response to mechanical stress. This stress induced autocrine response is suppressed by AT1R antagonists (Sadoshima et al., 1993) (Sadoshima and Izumo, 1993). However, subsequent studies identified also a stretch induced mechanostimulation of the AT1R leading to an AII independent activation of ERKs, in what was the first observation of mechanosensitivity of a GPCR (Zou et al., 2004). Mechanical stretch did not activate ERKs in HEK293 cells or COS7 cells, cells endogenously devoid of AT1R; however heterologous expression of the AT1R gave these cells the ability to respond to stretch (**Figure 1a**). On the other hand, cells transfected with the β2-AR did not display any similar level of mechanical activation (**Figure 1b**). Activated AT1R upon stretching interacts with the heterotrimeric G protein, resulting in Gαq11 redistribution in the cytosol and this was inhibited by AT1R blocker candesartan. Three key cardiovascular pathophysiological processes that have been identified are: 1) cardiomyocyte hypertrophy (Zou et al., 2004) (Zhou et al., 2010). At the molecular level this is mediated by $G_q$ activation leading to both $Ca^{2+}$ influx in the cytosol, as well as to Protein Kinase C (PKC) activation (Rosendorff, 1996); 2) Myogenic vasoconstriction (y Schnitzler et al., 2008), elicited at the molecular level by PKC-mediated activation of transient receptor potential channels (TRPCs); 3) abdominal aortic aneurism, mediated by β-arrestin/ERK1/2 signaling in vascular smooth muscle cells (Hall et al., 2021).

Mechanical stress also activated the AT1R in pressure overload induced cardiac hypertrophy mice model without the involvement of AII (Zou et al., 2004) .

In order to elucidate the conformational changes taking place at the AT1R under mechanical stress, the receptor was investigated using a Substituted Cysteine Accessibility Method assay, which allowed to track the relative conformational changes of the receptor. It was found that mechanical stretch increases the accessibility of Cys289 by inducing anticlockwise rotation of TM7. The inverse agonist, candesartan suppressed this stretch-induced activation of AT1R. With candesartan treatment the clockwise rotations of TM6 and TM7 shift the receptor to an inactive state (Yasuda et al., 2008).

Another factor that was investigated as a source of mechanoactivation for the AT1R is the change of the local lipid bilayer thickness. Molecular dynamics simulations of AT1R were performed incorporating the AT1R into equilibrated membrane patches consisting of 50 lipids per leaflet, maintained under physiological conditions of 37°C and 1 atm pressure. The conformational changes of the AT1R as a function of the change in membrane thickness was monitored by MD of the TM1-TM6 and TM3-TM6 distances. TM6 is a major player in binding of effector G proteins or β-arrestins. It was found that large thickness 1-stearoyl-2-oleoyl-sn-glycero-3-phosphocholine (SOPC) membrane increased the values of TM1-TM6 distances favoring the outward movement of TM6. The thinner dimyristoyl-sn-glycero-3-phosphocholine (DMPC) and 1-palmitoyl-2-oleoyl-glycero-3-phosphocholine (POPC) membranes present a greater hindrance to the mobility of TM6, as these membranes would require a higher local deformation than SOPC. Hence, membrane deformations near the AT1R cause a hydrophobic mismatch which plays an important role in the movement of TM6 which could activate or stabilize conformational changes that could facilitate downstream signaling (Poudel and Vanegas, 2024).

Further to these findings, Rakesh *et al.* found that, in cells and ex vivo heart preparations, mechanical stress activated a signaling pathway that required neither angiotensin II nor G proteins. Instead, β-arrestin was recruited to AT1R, the complex was internalized, and β-arrestin mediated activation of an antiapoptotic signaling pathway through ERK and Akt unfolded. Treating mice with the angiotensin receptor blocker losartan led to increased cardiomyocyte apoptosis, leading the authors to suggest that these drugs may block β-arrestin–mediated protective signaling in response to mechanical stress (Rakesh et al., 2010).

To better investigate this process, Tang et al generated an AT1R-β-arrestin2 fusion protein. Upon osmotic swelling (143 mOsm) the AT1R-β-arrestin2 fusion protein displayed increased ligand binding when cells were stimulated with the biased agonist TRV120023 (Figure 2). Membrane stretch preferentially favours the biased agonist-dependent ERK1/2 phosphorylation. BRET measurements displayed that after the osmotic swelling, the AT1R is allosterically biased into a β-arrestin2 activating conformation which cannot be changed into a different stable conformation even with addition of AngII. (Tang et al., 2014)). It shall be noted that in these experiments the authors observed also a ligand independent receptor activation, as reported earlier (Zou et al., 2004).

The AT1R thus embodies both pathways of mechanosensitive GPCR response, namely modulation of ligand-dependent activation as well as ligand-independent, constitutive activation. Up to 2023, it was the only non-adhesion mechanosensitive GPCR to display this behavior, as opposed to the other non-adhesion mechanosensitive GPCRs like bradykynin 2 (Chachisvilis et al., 2006), PTHR1(Zhang et al., 2009), 5-HT1R (Candelario and Chachisvilis, 2012), H1R(Erdogmus et al., 2019), formyl peptide receptor(Makino et al., 2006), apelin receptor (Scimia et al., 2012), GPR68(Xu et al., 2018) that have been observed to display mainly a ligand independent activation by mechanical stimulation.

**Mechano-response of the β2-AR**

Two reports this year have however highlighted this dual mechanism in another class A GPCR, also of notable cardiovascular relevance, namely the β2-AR. The β2-AR is regarded as a canonically $G_{\alpha s}$ coupled receptor, and its stimulation leads to increases in the intracellular cAMP levels. Activated β2-AR are then phosphorylated by GPCR Kinases, and signaling at the plasma membrane is terminated by recruitment of β-arrestins. However, β-arrestins are also considered bona-fide signal transducers, leading to activation of MAPK/ERK pathway. Additionally, there is evidence of the ability of the β2-AR to switch to signaling via the inhibitory G protein $G_i$(Nibley and Shenoy, 2024), albeit the mechanisms leading to this switch in cardiomyocytes are still being actively investigated (Strohman et al., 2019).

The β2-AR is one of the first GPCRs to be studied and has provided the blueprint for interpretation of behavior of most class A GPCRs. In this context, any observation of mechanosensitivity is clearly of great relevance. Cullum et al have reported its ligand-independent activation upon mechanical stress (Cullum et al., 2024), whereas Sirbu et al. observed the ligand-dependent regulation of its downstream response upon osmotic swelling (Sirbu et al., 2024b).

Cullum et al. investigated the β2-AR mediated cAMP responses using a luminescence-based biosensor assay: upon receptor activation, cAMP induces conformational alterations in the cAMP-binding domain of the protein kinase A (PKA) regulatory subunit (RIIβB), which are quantified by the change in luminescence of the biosensor. The basal cAMP levels of both endogenous and overexpressed β2-AR cell lines were similar. However, variations were observed when the plate hosting HEK293 recombinant TS-SNAP-β2AR cell line was removed and reinserted into the plate reader, suggesting that mechanical stimulation induced the changes in cAMP levels (**Figure 3**). Sequential mechano-stimulation of the receptor in HEK293G-β2AR by movement of the plate in and out of the plate-reader resulted in rapid increase in baseline luminescence. This was inhibited by the addition of the inverse agonist ICI-118551.

Notably the peak response to the vehicle HBSS was significantly higher in HEK293G overexpressing β2-AR than in the cell line displaying endogenous β2-AR expression (Figure 3d) and the difference of luminescence values taken after the initial plate loading with respect to basal luminescence readings in HEK293G was non-significant (Figure 3d). Of interest, four β2-AR inverse agonists- ICI-118551, carvedilol, carazolol, propranolol added by removing the plate and re-inserting into the plate reader reversed the initial basal cAMP responses. These results suggest that mechanical stimulation enhances the constitutive R* active state of the receptor, as confirmed by the fact that altering the residue D113 which is the catecholamine binding domain of β2-AR, reduced the basal cAMP responses upon mechanical stimulation and reduced the affinity of ICI-118551. Overall, the study by Cullum et al highlights a mechanical activation of the β2-AR, that can be countered by the use of an inverse agonist for this receptor, suggesting for the mechanical stress/force a role akin to an orthosteric ligand. We shall however note that, as expected, the magnitude of the cAMP response is dependent on the overall β2-AR (over)-expression of the cells, raising questions about the physiological significance of this specific mechanism.

This result stands in contrast with previous work where the role of extracellular β2-AR-associated glycan chains has shown to play an important role in mechanical stimulation of β2-AR in the form of an allosteric modulation of the receptor signaling along its β-Arrestin - ERK activation pathway. Notably, no significant increase in cAMP levels was observed in cells incubated with meningococcal type IV pili, shown to interact with N-terminal receptor glycosylation domains (Figure 4), as opposed to substantial recruitment of β-Arrestin2 (Coureuil et al., 2010).

In a follow up study by the same group, Virion et al investigated the role of receptor glycosylation by analysing mutants in which asparagine residues were substituted with alanine. These mutations occurred at the N-terminal (N6A and N15A) and in the second extracellular loop (N187A) of the β2-AR. Mutants showed an attenuated basal cAMP response to the mechanical stimulation by bacterial pili and were also insensitive to inverse agonist (prop) treatment (Virion et al., 2019). In summary, this set of studies observed a ligand independent β-arrestin mechanoactivation mediated by the β2-AR, but no orthosteric activation of the receptor along the Gs-cAMP axis by such mechanical stimulation.

Part of these observations were recapitulated in a recent study by Sirbu et al. The authors employed osmotic swelling (also referred to as hypotonic shock) as a probe to alter the local biophysical environment of endogenous adrenoreceptors. Most of the cells in the body are highly sensitive to osmotic stress which leads to a diverse range of mechanisms like rearrangement of the cell membrane components, modulation of the actin cytoskeleton and overall increase in the cell area (Hoffmann et al., 2009, Groulx et al., 2007). We shall note here that osmotic swelling has been routinely used as a proxy for causing cellular stretching, including in several of the studies referred to in this review. Nonetheless, both the strength of the treatment and the time of measurement from the onset of the swelling vary largely between different laboratories (Hoffmann et al., 2009, Groulx et al., 2007). Sirbu et al used values in line with those that demonstrated previouslz ligand-independent Gq activation(Erdogmus et al., 2019). To study the effect of osmotic swelling on Gαs coupled β2-AR, the cAMP levels were monitored first in HEK293 cells using a FRET-based cAMP fluorescence biosensor Epac-SH187. The kinetic cAMP response to 200mOsm hypotonic solution as compared to 300mOsm isotonic solution in presence of the agonist isoproterenol was observed to reach a significantly higher peak in HEK293 cells, thereby subsiding to a steady state level comparable between hypotonic and isotonic conditions. After ruling out changes to the adenyl cyclase activation, potential inhibition of PDE activity as well as effects related to altered receptor internalization Gαs was suspected to be the key player in mediating the response. This possibility was investigated by measuring the recruitment of nanobody 37 to the cell's plasma membrane which binds to the nucleotide-free Gαs in the agonist-bound β2-AR Gαs signaling complex. Osmotic swelling of cells co-transfected with β2-AR and Gs was found to be beneficial in the Iso - β2-AR - Gαs ternary complex formation as the Nb37 recruitment was almost double as compared to the non-swollen cells. It shall be noted that β-Arrestin2 recruitment was also increased by almost two-fold in swollen cells and it was found to be independent of Gαs activation. Next, fluorescent anisotropy studies revealed the affinity towards the agonist was significantly higher in swollen cells. These findings confirm that osmotic swelling promotes active β2-AR conformations favorable both to Gαs and and β-Arrestin2 recruitment and ultimately leads to an increased cAMP level (Sirbu et al., 2024b).

This mechanism was further observed in swollen adult cardiomyocytes isolated from transgenic mouse expressing the Epac1-camps cAMP sensor (Figure 5).

Cardiomyocytes are known to physiologically experience swelling during reperfusion after ischemic shock; the concomitant observation that the chronotropic response upon swelling in human induced pluripotent stem cells-derived cardiomyocytes (hiPSC-CMs) was also increased (this is consistent with improved Ca2+ handling caused by a downstream activation of PKA in response to the transient increase in cAMP levels in swollen cells), upholds the relevance of this mechanoactivation of the β2-AR in cardiac physiology.

Taken together, these three studies on the mechanical sensitivity of the β2-AR paint a complex picture of the activation landscape of this receptor, and of how this can be modulated by mechanical cues. Sirbu et al, Coureuil et al & Virion et al point to an allosteric modulatory effect of the mechanical stimuli at the receptor. Moreover, these works also highlight that mechanical stimulation clearly elicits a β-arrestin dependent pathway. Somewhat differently to what is observed by Cullum et al, β-blockade does not abolish the effect of hypotonic shock on receptor Gs-mediated signaling; instead, in osmotically swollen cells, it is easier for orthosteric agonists to break the blockade (Figure 6).

Prominently, membrane stretch caused by osmotic swelling does not cause any increase in receptors constitutive activity, when measured by downstream cAMP readouts, in line with the action of the bacterial pili on the N-terminal glycans of the β2-AR, which does not cause an increase in basal cAMP level. On the other hand, the mechanical stimulation by linear movement of the plate in and out of a plate reader was sufficient to cause substantial increases in cAMP, as observed by Cullum et al, which in turn appears to be associated to a fully active conformation of the receptor, since inverse agonists are able to shift the equilibrium back from the active to the inactive state. We shall also note here that in the study by Cullum et al, while the effects are most prominent in case of receptor overexpression, they approach the limit of statistical significance when receptors are endogenously expressed. Moreover, it was not investigated whether the mechanical loading of the plate also triggers the activation of β-Arrestins to check if there is also biased activation of the β2-AR.

It is possible that mechanical stimulation leads to different conformational outcomes at the β2-AR depending on whether the stimulus is enacted via extracellular domains (Virion et al., 2019) and associated glycans, or by intracellularly driven changes in osmotic pressure, membrane curvature and actin remodeling.

As observed in the β1-AR, the application of pressure shifts the equilibrium of the agonist-bound receptor towards a fully active conformation. This was a result of the intracellular G protein binding pocket widening in the fully active state to accommodate helix α5 of the G protein. This increased hydration of the G protein binding pocket due to the reduction in the void volume was shown to enhance the agonist affinity by approximately 100- fold (Abiko et al., 2019).

**Summary of AT1R and β2-AR mechanotransduction: molecular mechanism and physiological**

**implications**

The results discussed in the sections above are graphically summarized in Figure 7, specifically highlighting the molecular pathways elicited by each instance of mechanical stimulation at either the β2-AR or the AT1R.

The ability of GPCRs to sense and respond to mechanical forces has profound implications for cardiovascular health and disease. Overall, the broader implications of such mechanosensitive behavior in the β2-AR are still unclear, as well as their role in maintaining vascular and cardiac health. In conditions such as hypertension and heart failure, the cardiovascular system experiences increased mechanical stress, which can alter GPCR signaling and contribute to disease progression. As discussed, mechanical activation of AT1R in response to elevated blood pressure can exacerbate vasoconstriction and hypertrophy, promoting adverse cardiovascular remodeling. We shall note here that while both the β2-AR and AT1R are described in this context as mechanosensitive, evidence of mechanosensitivity for the AT1R dates back to 2004(Zou et al., 2004), allowing to build-up a more robust collection of pathophysiological outcomes in a cardiovascular setting, whereas some of the most compelling data for the β2-AR arose just last year.

In the case of mechanical stimulation of the β2-AR, depending on the cause of mechanical stimulation, we can speculate that we would observe a potentiation of catecholamine effects in cardiomyocyte cells undergoing increasing stretch and/or osmotic swelling, which in the long term would contribute to adverse remodelling. Whereas mechanical β2-AR activation in the vascular system could lead to vasodilation and act as a positive feedback loop to relieve pressure. Whether physiologically observed mechanical stressors would lead to the effect of constitutive activation of the β2-AR observed for linear shaking of cells, as observed by Cullum et al. (Cullum et al., 2024), remains to be further investigated.

Overall ,understanding how mechanical stimuli influence GPCR function in these contexts could provide new insights into disease mechanisms and reveal novel therapeutic targets. Additionally, the concept of biased signaling, whereby mechanical forces preferentially activate specific GPCR signaling pathways, opens up exciting possibilities for the development of biased agonists that selectively target beneficial pathways while avoiding harmful ones.

**Mechanisms and novel methodological avenues**

Part of the reason why, even for a single receptor, a relative diversity of mechanosensitive behaviors has been observed, may hinge on the fact that application of a defined, controllable and localized mechanical stimulus remains challenging. Moreover, it is not yet clear which residues are critical for mediating mechanosensitive responses at selected receptors, hampering the ability to generate negative controls.

The aforementioned studies on β2-AR do not fall into this category. To better understand the molecular mechanisms of mechanosensitive responses, such as in the case of β2-AR, it is crucial to employ precise and controllable approaches to apply mechanical stimuli combined with techniques that allow for quantitative measurements at the single cell level. Novel microscopic techniques facilitate the studies on GPCR mechanosensitivity. Single-molecule fluorescence can track individual GPCRs in live cells, revealing changes to their dynamics (Scarselli et al., 2015). By providing precise measurements at the single-molecule level, single molecule force spectroscopy tools like Atomic Force Microscopy (AFM) or optical tweezers offer insights into the fundamental forces and dynamics that govern biological processes (Zocher et al., 2013).

Force spectroscopy approaches, such as optical tweezer (OT) technologies, coupled to microscopy, have been employed over the last three decades to study various biological phenomena but seldom to GPCR-related research. Since the invention of optical tweezers by Arthur Ashkin in 1970 (Ashkin and Dziedzic, 1971), OTs have been used to study many biological processes and also to study the effect of mechanical forces on cells (Ashkin and Dziedzic, 1989). Compared to other mechanical spectroscopy techniques such as atomic force microscopy (AFM) and magnetic tweezers, OT provide high-resolution position and force measurements in the piconewton (pN) range, making them ideal for studying the dynamic properties of cells (Arbore et al., 2019). OTs have enabled the study of red blood cell deformations, which is essential for linking the cell's mechanical properties like viscoelasticity, bending modulus and stretching force to diseases such as malaria and sickle cell anemia (Lim et al., 2004). Tethers, i.e. thin membrane tubules pulled between a microbead and the bulk of the cell, allowed to establish an important connection between membrane tension and actin cytoskeletal remodeling. Tethers pulled from NIH3T3 cells revealed the presence of actin within, which indicated that the cell responds to the force application by reorganization of its cytoskeletal filaments (Pontes et al., 2011). More recent studies, using a dual trapping geometry (De Belly et al., 2023) showed that using tethers pulled from only the membrane of cells (e.g. from blebs), the overall actin cortex opposes membrane tension propagation but forces that engage the actin cortex cause rapid long-range actin driven membrane tension propagation.

In the context of GPCR studies, tether pulling has been used to monitor selected GPCRs redistribution. The neuropeptide Y2 receptor (Y2R), was expressed transiently in HEK293 cells and the receptor distribution was measured by means of fluorescence intensity as a function of tether radius. The receptor sorting was inversely proportional to the radius of the tether. Follow up experiments in spontaneous PC12 cells filopodia displayed that agonist bound β1-AR, β2-AR and Y2R have different preference for the curved membrane of the tethers, i.e. agonist addition decreased the intrinsic curvature and increased the bending rigidity for Y2R and β2-AR which could be a mechanism for stabilizing the extracellular ligand binding pocket of the receptor. This seminal work thus suggested a role of membrane curvature as a regulator of GPCR activation and signalling (Rosholm et al., 2017). These data further matched confocal microscopy mapping of the basolateral membrane of β1-AR-transfected HEK293 cells, indicating a strong correlation between receptor density and the membrane's mean curvature. The lipid composition and interleaflet asymmetry of the membrane enhances receptor coupling to shallow membranes; however, shallow curvature alone is sufficient to establish GPCR-enriched and GPCR-depleted domains. An unambiguous curvature-dependent sorting of GPCRs in shallow membranes was observed for β1-AR, β2-AR, Y2R and glucagon-like peptide 1 receptor (GLP1R). Ligand induction shifted the distribution of GLP1R from receptor-depleted to receptor-enriched domains at regions of negative membrane curvature, likely due to a conformational change in the receptor (Kockelkoren et al., 2023). Thus, a strong correlation exists between membrane curvature and the spatial distribution of GPCR density, again indicating a geometry sensing feature of these receptors that could suggestively be connected to the above-reported mechanosensitive features; however, the mechanisms responsible for the formation of these high-density domains is still under investigation (Kockelkoren et al., 2023).

We argue that combining tether pulling experiments using optical tweezers to dynamic measurements of receptor diffusion as well as activation, e.g. using fluorescence biosensors, is a key avenue to unravel several of the challenges outlined in this review to address the mechanosensitive nature of these important receptors. Three important questions that can be addressed using this technology are: 1) the molecular determinants of receptor rearrangement in the curved compartments, which is important towards better understanding receptor compartmentalization processes at the membrane, e.g. β2-AR segregation in the t-tubules of adult cardiomyocytes (Bathe-Peters et al., 2021). 2) Decoupling the role of curvature and cytoskeletal interaction in affecting GPCR dynamics and trafficking. 3) Determine whether the downstream signaling cascade in regions that are mechanically deformed changes with respect to areas of the cell that are mechanically at rest.

These ideas are graphically summarised in Figure 8. Here, OT combined with confocal microscopy are used to pull tethers from HEK293 cells transfected with a fluorescently tagged membrane receptor (Figure 8A). Once a tether is established, it is possible to conduct a confocal linescan across the tether to generate a kymograph (Figure 7B). Then, fluorescence recovery after photobleaching(Edidin et al., 1976) can be used to monitor the diffusion rate of the receptors in the tether, and compare their diffusion to that observed on the basal membrane (Figure 8C). If the cells are additionally transfected with a biosensor for downstream second messenger, such as cAMP, one could monitor membrane conformation-dependent changes to downstream signaling (Figure 8D). In contrast to the earlier study by Rosholm et al (Rosholm et al., 2017), which calculated receptor density based on tether radius, kymographs offer dynamic information on local receptor abundance and its time dependence.

These approaches could then synergise with specific models tailored to study mechanostimulation. While there are not yet animal models that could be specifically used to monitor role of mechanosensitivity, Erdogmus et al. (Erdogmus et al., 2019) used Gq specific knockouts to check for the effect of mechanical activation of Gq-coupled GPCRs such as H1R. Ex vivo mutants of the H1R lacking helix 8, the amphiphatic alpha-helical domain initiating just after the conserved NPXXY motif in TM7, were shown to lose mechano transduction. We have not identified any specific animal model carrying these mutations, but we can speculate here that mutants disrupting H8 alpha-helical structure, such as exchanging phenylalanine for prolines or, potentially, affecting palmitoylation sites, would serve the purpose. A list of class A GPCR alignments indicating Helix 8 residues (Huynh et al., 2009) could be the starting point to identify mutations in relevant animal models or gene-edited hiPS cell lines.

We believe that these new approaches will help shed further light into the mechanosensitive behavior of receptors and how this affects their subcellular downstream signaling profiles, with generalisation possible beyond the cardiovascular field.

**Funding**

PA gratefully acknowledges support from the Leverhulme Trust (RL-2022-015) and from the UKRI BBSRC

(BB/X019047/1).

## References

ABIKO, L. A., GRAHL, A. & GRZESIEK, S. 2019. High Pressure Shifts the β1-Adrenergic Receptor to the Active Conformation in the Absence of G Protein. *Journal of the American Chemical Society,* 141**,** 16663-16670.
ARBORE, C., PEREGO, L., SERGIDES, M. & CAPITANIO, M. 2019. Probing force in living cells with optical tweezers: from single-molecule mechanics to cell mechanotransduction. *Biophysical Reviews,* 11**,** 765-782.
ASHKIN, A. & DZIEDZIC, J. M. 1971. Optical Levitation by Radiation Pressure. *Applied Physics Letters,* 19**,** 283-285.
ASHKIN, A. & DZIEDZIC, J. M. 1989. Internal cell manipulation using infrared laser traps. *Proceedings of the National Academy of Sciences,* 86**,** 7914-7918.
BACCOUCH, R., RASCOL, E., STOKLOSA, K. & ALVES, I. D. 2022. The role of the lipid environment in the activity of G protein coupled receptors. *Biophys Chem,* 285**,** 106794.
BATHE-PETERS, M., GMACH, P., BOLTZ, H.-H., EINSIEDEL, J., GOTTHARDT, M., HÜBNER, H., GMEINER, P., LOHSE, M. J. & ANNIBALE, P. 2021. Visualization of β-adrenergic receptor dynamics and differential localization in cardiomyocytes. *Proceedings of the National Academy of Sciences,* 118.
CANDELARIO, J. & CHACHISVILIS, M. 2012. Mechanical Stress Stimulates Conformational Changes in 5-Hydroxytryptamine Receptor 1B in Bone Cells. *Cellular and Molecular Bioengineering,* 5**,** 277-286.
CHACHISVILIS, M., ZHANG, Y.-L. & FRANGOS, J. A. 2006. G protein-coupled receptors sense fluid shear stress in endothelial cells. *Proceedings of the National Academy of Sciences,* 103**,** 15463-15468.
COUREUIL, M., LÉCUYER, H., SCOTT, M. G. H., BOULARAN, C., ENSLEN, H., SOYER, M., MIKATY, G., BOURDOULOUS, S., NASSIF, X. & MARULLO, S. 2010. Meningococcus Hijacks a β2-Adrenoceptor/β-Arrestin Pathway to Cross Brain Microvasculature Endothelium. *Cell,* 143**,** 1149-1160.
CULLUM, S. A., PLATT, S., DALE, N., ISAAC, O. C., WRAGG, E. S., SOAVE, M., VEPRINTSEV, D. B., WOOLARD, J., KILPATRICK, L. E. & HILL, S. J. 2024. Mechano-sensitivity of β2-adrenoceptors enhances constitutive activation of cAMP generation that is inhibited by inverse agonists. *Communications Biology,* 7.
DAMIAN, M., LOUET, M., GOMES, A. A. S., M'KADMI, C., DENOYELLE, S., CANTEL, S., MARY, S., BISCH, P. M., FEHRENTZ, J. A., CATOIRE, L. J., FLOQUET, N. & BANERES, J. L. 2021. Allosteric modulation of ghrelin receptor signaling by lipids. *Nat Commun,* 12**,** 3938.
DAWALIBY, R., TRUBBIA, C., DELPORTE, C., MASUREEL, M., VAN ANTWERPEN, P., KOBILKA, B. K. & GOVAERTS, C. 2016. Allosteric regulation of G protein-coupled receptor activity by phospholipids. *Nat Chem Biol,* 12**,** 35-9.
DE BELLY, H., YAN, S., BORJA DA ROCHA, H., ICHBIAH, S., TOWN, J. P., ZAGER, P. J., ESTRADA, D. C., MEYER, K., TURLIER, H., BUSTAMANTE, C. & WEINER, O. D. 2023. Cell protrusions and contractions generate long-range membrane tension propagation. *Cell,* 186**,** 3049-3061.e15.
EDIDIN, M., ZAGYANSKY, Y. & LARDNER, T. J. 1976. Measurement of Membrane Protein Lateral Diffusion in Single Cells. *Science,* 191**,** 466-468.
ERDOGMUS, S., STORCH, U., DANNER, L., BECKER, J., WINTER, M., ZIEGLER, N., WIRTH, A., OFFERMANNS, S., HOFFMANN, C., GUDERMANN, T. & MEDEROS Y SCHNITZLER, M. 2019. Helix 8 is the essential structural motif of mechanosensitive GPCRs. *Nature Communications,* 10.
FRIEDRICH, O., WAGNER, S., BATTLE, A. R., SCHÜRMANN, S. & MARTINAC, B. 2012. Mechano-regulation of the beating heart at the cellular level – Mechanosensitive channels in normal and diseased heart. *Progress in Biophysics and Molecular Biology,* 110**,** 226-238.
GAROFFOLO, G. & PESCE, M. 2019. Mechanotransduction in the Cardiovascular System: From Developmental Origins to Homeostasis and Pathology. *Cells,* 8.

GETHER, U. 2000. Uncovering Molecular Mechanisms Involved in Activation of G Protein-Coupled Receptors. *Endocrine Reviews,* 21**,** 90-113.

GHANOUNI, P., STEENHUIS, J. J., FARRENS, D. L. & KOBILKA, B. K. 2001. Agonist-induced conformational changes in the G-protein-coupling domain of the $\beta_2$ adrenergic receptor. *Proceedings of the National Academy of Sciences,* 98**,** 5997-6002.

GROULX, N., BOUDREAULT, F., ORLOV, S. N. & GRYGORCZYK, R. 2007. Membrane Reserves and Hypotonic Cell Swelling. *Journal of Membrane Biology,* 214**,** 43-56.

GUNSEL, U. & HAGN, F. 2022. Lipid Nanodiscs for High-Resolution NMR Studies of Membrane Proteins. *Chem Rev,* 122**,** 9395-9421.

HALL, S., WARD, N. D., PATEL, R., AMIN-JAVAHERI, A., LANFORD, H., GRESPIN, R. T., COUCH, C., XIONG, Y., MUKHERJEE, R., JONES, J. A. & RUDDY, J. M. 2021. Mechanical activation of the angiotensin II type 1 receptor contributes to abdominal aortic aneurysm formation. *JVS-Vascular Science,* 2**,** 194-206.

HARDMAN, K., GOLDMAN, A. & PLIOTAS, C. 2023. Membrane force reception: mechanosensation in G protein-coupled receptors and tools to address it. *Current Opinion in Physiology,* 35.

HAUSER, A. S., ATTWOOD, M. M., RASK-ANDERSEN, M., SCHIÖTH, H. B. & GLORIAM, D. E. 2017. Trends in GPCR drug discovery: new agents, targets and indications. *Nature Reviews Drug Discovery,* 16**,** 829-842.

HAUSER, A. S., KOOISTRA, A. J., MUNK, C., HEYDENREICH, F. M., VEPRINTSEV, D. B., BOUVIER, M., BABU, M. M. & GLORIAM, D. E. 2021. GPCR activation mechanisms across classes and macro/microscales. *Nature Structural & Molecular Biology,* 28**,** 879-888.

HILGER, D., MASUREEL, M. & KOBILKA, B. K. 2018. Structure and dynamics of GPCR signaling complexes. *Nat Struct Mol Biol,* 25**,** 4-12.

HOFFMANN, E. K., LAMBERT, I. H. & PEDERSEN, S. F. 2009. Physiology of Cell Volume Regulation in Vertebrates. *Physiological Reviews,* 89**,** 193-277.

HOLLOWAY, A. C., QIAN, H., PIPOLO, L., ZIOGAS, J., MIURA, S.-I., KARNIK, S., SOUTHWELL, B. R., LEW, M. J. & THOMAS, W. G. 2002. Side-Chain Substitutions within Angiotensin II Reveal Different Requirements for Signaling, Internalization, and Phosphorylation of Type 1A Angiotensin Receptors. *Molecular Pharmacology,* 61**,** 768-777.

HUYNH, J., THOMAS, W. G., AGUILAR, M.-I. & PATTENDEN, L. K. 2009. Role of helix 8 in G protein-coupled receptors based on structure–function studies on the type 1 angiotensin receptor. *Molecular and Cellular Endocrinology,* 302**,** 118-127.

INGBER, D. E. 2006. Cellular mechanotransduction: putting all the pieces together again. *The FASEB Journal,* 20**,** 811-827.

JIANG, H., GALTES, D., WANG, J. & ROCKMAN, H. A. 2022. G protein-coupled receptor signaling: transducers and effectors. *American Journal of Physiology-Cell Physiology,* 323**,** C731-C748.

JONES, A. J. Y., GABRIEL, F., TANDALE, A. & NIETLISPACH, D. 2020. Structure and Dynamics of GPCRs in Lipid Membranes: Physical Principles and Experimental Approaches. *Molecules,* 25.

KOCKELKOREN, G., LAURITSEN, L., SHUTTLE, C. G., KAZEPIDOU, E., VONKOVA, I., ZHANG, Y., BREUER, A., KENNARD, C., BRUNETTI, R. M., D'ESTE, E., WEINER, O. D., ULINE, M. & STAMOU, D. 2023. Molecular mechanism of GPCR spatial organization at the plasma membrane. *Nature Chemical Biology,* 20**,** 142-150.

LEFKOWITZ, R. J. 2013. A Brief History of G-Protein Coupled Receptors (Nobel Lecture). *Angewandte Chemie International Edition,* 52**,** 6366-6378.

LIM, C. T., DAO, M., SURESH, S., SOW, C. H. & CHEW, K. T. 2004. Large deformation of living cells using laser traps. *Acta Materialia,* 52**,** 1837-1845.

MAHONEY, J. P. & SUNAHARA, R. K. 2016. Mechanistic insights into GPCR–G protein interactions. *Current Opinion in Structural Biology,* 41**,** 247-254.

MAKINO, A., PROSSNITZ, E. R., BÜNEMANN, M., WANG, J. M., YAO, W. & SCHMID-SCHÖNBEIN, G. W. 2006. G protein-coupled receptors serve as mechanosensors for fluid shear stress in neutrophils. *American Journal of Physiology-Cell Physiology,* 290**,** C1633-C1639.

MANGLIK, A. & KOBILKA, B. 2014. The role of protein dynamics in GPCR function: insights from the β2AR and rhodopsin. *Current Opinion in Cell Biology,* 27**,** 136-143.

MANGLIK, A. & KRUSE, A. C. 2017. Structural Basis for G Protein-Coupled Receptor Activation. *Biochemistry,* 56**,** 5628-5634.

MANNA, M., NIEMELA, M., TYNKKYNEN, J., JAVANAINEN, M., KULIG, W., MULLER, D. J., ROG, T. & VATTULAINEN, I. 2016. Mechanism of allosteric regulation of beta(2)-adrenergic receptor by cholesterol. *Elife,* 5.

MAUDSLEY, S., MARTIN, B. & LUTTRELL, L. M. 2005. The origins of diversity and specificity in g protein-coupled receptor signaling. *J Pharmacol Exp Ther,* 314**,** 485-94.

NIBLEY, P. C. & SHENOY, S. K. 2024. β-adrenergic receptor signaling mediated by β-arrestins and its potential role in heart failure. *Current Opinion in Physiology,* 37.

PANI, B., AHN, S., RAMBARAT, P. K., VEGE, S., KAHSAI, A. W., LIU, A., VALAN, B. N., STAUS, D. P., COSTA, T. & LEFKOWITZ, R. J. 2021. Unique Positive Cooperativity Between the β-Arrestin–Biased β-Blocker Carvedilol and a Small Molecule Positive Allosteric Modulator of the β2-Adrenergic Receptor. *Molecular Pharmacology,* 100**,** 513-525.

PERSECHINO, M., HEDDERICH, J. B., KOLB, P. & HILGER, D. 2022. Allosteric modulation of GPCRs: From structural insights to in silico drug discovery. *Pharmacology & Therapeutics,* 237.

PONTES, B., VIANA, N. B., SALGADO, L. T., FARINA, M., NETO, V. M. & NUSSENZVEIG, H. M. 2011. Cell Cytoskeleton and Tether Extraction. *Biophysical Journal,* 101**,** 43-52.

POUDEL, B. & VANEGAS, J. M. 2024. Structural Rearrangement of the AT1 Receptor Modulated by Membrane Thickness and Tension. *The Journal of Physical Chemistry B,* 128**,** 9470-9481.

RAKESH, K., YOO, B., KIM, I.-M., SALAZAR, N., KIM, K.-S. & ROCKMAN, H. A. 2010. β-Arrestin–Biased Agonism of the Angiotensin Receptor Induced by Mechanical Stress. *Science Signaling,* 3.

ROG, T., PASENKIEWICZ-GIERULA, M., VATTULAINEN, I. & KARTTUNEN, M. 2009. Ordering effects of cholesterol and its analogues. *Biochim Biophys Acta,* 1788**,** 97-121.

ROSENDORFF, C. 1996. The Renin-Angiotensin System and Vascular Hypertrophy. *Journal of the American College of Cardiology,* 28**,** 803-812.

ROSHOLM, K. R., LEIJNSE, N., MANTSIOU, A., TKACH, V., PEDERSEN, S. L., WIRTH, V. F., ODDERSHEDE, L. B., JENSEN, K. J., MARTINEZ, K. L., HATZAKIS, N. S., BENDIX, P. M., CALLAN-JONES, A. & STAMOU, D. 2017. Membrane curvature regulates ligand-specific membrane sorting of GPCRs in living cells. *Nature Chemical Biology,* 13**,** 724-729.

SADOSHIMA, J. & IZUMO, S. 1993. Mechanical stretch rapidly activates multiple signal transduction pathways in cardiac myocytes: potential involvement of an autocrine/paracrine mechanism. *The EMBO Journal,* 12**,** 1681-1692.

SADOSHIMA, J.-I., XU, Y., SLAYTER, H. S. & IZUMO, S. 1993. Autocrine release of angiotensin II mediates stretch-induced hypertrophy of cardiac myocytes in vitro. *Cell,* 75**,** 977-984.

SCARSELLI, M., ANNIBALE, P., MCCORMICK, P. J., KOLACHALAM, S., ARINGHIERI, S., RADENOVIC, A., CORSINI, G. U. & MAGGIO, R. 2015. Revealing G-protein-coupled receptor oligomerization at the single-molecule level through a nanoscopic lens: methods, dynamics and biological function. *The FEBS Journal,* 283**,** 1197-1217.

SCIMIA, M. C., HURTADO, C., RAY, S., METZLER, S., WEI, K., WANG, J., WOODS, C. E., PURCELL, N. H., CATALUCCI, D., AKASAKA, T., BUENO, O. F., VLASUK, G. P., KALIMAN, P., BODMER, R., SMITH, L. H., ASHLEY, E., MERCOLA, M., BROWN, J. H. & RUIZ-LOZANO, P. 2012. APJ acts as a dual receptor in cardiac hypertrophy. *Nature,* 488**,** 394-398.

SHYY, J. Y.-J. & CHIEN, S. 2002. Role of Integrins in Endothelial Mechanosensing of Shear Stress. *Circulation*

*Research,* 91**,** 769-775.
SIRBU, A., BATHE-PETERS, M., KUMAR, J. L. M., INOUE, A., LOHSE, M. J. & ANNIBALE, P. 2024a. Cell swelling enhances ligand-driven beta-adrenergic signaling. *Nat Commun,* 15**,** 7822.
SIRBU, A., BATHE-PETERS, M., KUMAR, J. L. M., INOUE, A., LOHSE, M. J. & ANNIBALE, P. 2024b. Cell swelling enhances ligand-driven β-adrenergic signaling. *Nature Communications,* 15.
STORCH, U., SCHNITZLER, M. M. Y. & GUDERMANN, T. 2012. G protein-mediated stretch reception. *American Journal of Physiology-Heart and Circulatory Physiology,* 302**,** H1241-H1249.
STROHMAN, M. J., MAEDA, S., HILGER, D., MASUREEL, M., DU, Y. & KOBILKA, B. K. 2019. Local membrane charge regulates β2 adrenergic receptor coupling to Gi3. *Nature Communications,* 10.
TAGHON, G. J., ROWE, J. B., KAPOLKA, N. J. & ISOM, D. G. 2021. Predictable cholesterol binding sites in GPCRs lack consensus motifs. *Structure,* 29**,** 499-506 e3.
TANG, W., STRACHAN, R. T., LEFKOWITZ, R. J. & ROCKMAN, H. A. 2014. Allosteric Modulation of β-Arrestin-biased Angiotensin II Type 1 Receptor Signaling by Membrane Stretch. *Journal of Biological Chemistry,* 289**,** 28271-28283.
VIRION, Z., DOLY, S., SAHA, K., LAMBERT, M., GUILLONNEAU, F., BIED, C., DUKE, R. M., RUDD, P. M., ROBBE-MASSELOT, C., NASSIF, X., COUREUIL, M. & MARULLO, S. 2019. Sialic acid mediated mechanical activation of β2 adrenergic receptors by bacterial pili. *Nature Communications,* 10.
WEIS, W. I. & KOBILKA, B. K. 2018. The Molecular Basis of G Protein–Coupled Receptor Activation. *Annual Review of Biochemistry,* 87**,** 897-919.
WILDE, C., MITGAU, J., SUCHÝ, T., SCHÖNEBERG, T. & LIEBSCHER, I. 2022. Translating the force—mechano-sensing GPCRs. *American Journal of Physiology-Cell Physiology,* 322**,** C1047-C1060.
XIAO, R., LIU, J. & SHAWN XU, X. Z. 2023. Mechanosensitive GPCRs and ion channels in shear stress sensing. *Current Opinion in Cell Biology,* 84.
XU, J., MATHUR, J., VESSIÈRES, E., HAMMACK, S., NONOMURA, K., FAVRE, J., GRIMAUD, L., PETRUS, M., FRANCISCO, A., LI, J., LEE, V., XIANG, F.-L., MAINQUIST, J. K., CAHALAN, S. M., ORTH, A. P., WALKER, J. R., MA, S., LUKACS, V., BORDONE, L., BANDELL, M., LAFFITTE, B., XU, Y., CHIEN, S., HENRION, D. & PATAPOUTIAN, A. 2018. GPR68 Senses Flow and Is Essential for Vascular Physiology. *Cell,* 173**,** 762-775.e16.
Y SCHNITZLER, M. M., STORCH, U., MEIBERS, S., NURWAKAGARI, P., BREIT, A., ESSIN, K., GOLLASCH, M. & GUDERMANN, T. 2008. Gq-coupled receptors as mechanosensors mediating myogenic vasoconstriction. *The EMBO Journal,* 27**,** 3092-3103.
YASUDA, N., MIURA, S. I., AKAZAWA, H., TANAKA, T., QIN, Y., KIYA, Y., IMAIZUMI, S., FUJINO, M., ITO, K., ZOU, Y., FUKUHARA, S., KUNIMOTO, S., FUKUZAKI, K., SATO, T., GE, J., MOCHIZUKI, N., NAKAYA, H., SAKU, K. & KOMURO, I. 2008. Conformational switch of angiotensin II type 1 receptor underlying mechanical stress-induced activation. *EMBO reports,* 9**,** 179-186.
ZHANG, Y.-L., FRANGOS, J. A. & CHACHISVILIS, M. 2009. Mechanical stimulus alters conformation of type 1 parathyroid hormone receptor in bone cells. *American Journal of Physiology-Cell Physiology,* 296**,** C1391-C1399.
ZHOU, N., LI, L., WU, J., GONG, H., NIU, Y., SUN, A., GE, J. & ZOU, Y. 2010. Mechanical stress-evoked but angiotensin II-independent activation of angiotensin II type 1 receptor induces cardiac hypertrophy through calcineurin pathway. *Biochemical and Biophysical Research Communications,* 397**,** 263-269.
ZOCHER, M., BIPPES, C. A., ZHANG, C. & MÜLLER, D. J. 2013. Single-molecule force spectroscopy of G-protein-coupled receptors. *Chemical Society Reviews,* 42.
ZOCHER, M., ZHANG, C., RASMUSSEN, S. G. F., KOBILKA, B. K. & MÜLLER, D. J. 2012. Cholesterol increases kinetic, energetic, and mechanical stability of the human β2-adrenergic receptor. *Proceedings of the National Academy of Sciences,* 109.
ZOU, Y., AKAZAWA, H., QIN, Y., SANO, M., TAKANO, H., MINAMINO, T., MAKITA, N., IWANAGA, K., ZHU, W., KUDOH, S., TOKO, H., TAMURA, K., KIHARA, M., NAGAI, T., FUKAMIZU, A.,

UMEMURA, S., IIRI, T., FUJITA, T. & KOMURO, I. 2004. Mechanical stress activates angiotensin II type 1 receptor without the involvement of angiotensin II. *Nature Cell Biology,* 6**,** 499-506.

## Figure Legends

**Figure 1: Evidence of mechanoactivation of the AT1R.** A) HEK293-AT1-WT cells were stretched in the absence (−) or presence of 100 nM (Sar1,Ile8)-AII. B) COS7 cells transiently transfected with β2-AR receptors were pretreated with the β-AR blocker propranolol (Prop) or vehicle (−) and then stimulated with mechanical stretch or Isoproterenol. Adapted with permission from (Zou et al., 2004).

**Figure 2: Allosteric modulation of hypotonic swelling on β-arrestin recruitment to the AT1R.** A), competition binding isotherms for the balanced agonist AngII. No curve shift was observed under hypotonic osmotic stretch (143 mOsm/kg). B), competition binding isotherms for the β-arrestin-biased ligand TRV120023 showing that it bound with 9.7-fold greater affinity to the AT1R-β-arrestin2 under isotonic condition, with a further 3.5-fold shift in binding affinity under hypotonic conditions. Adapted from (Tang et al., 2014).

**Figure 3: mechanical activation of β2-AR, dependence on overexpression.** A) Impact of repeated mechanical stimulation on the basal GloSensorTM time-course responses in a clonal HEK293G cell line overexpressing recombinant TS- SNAP-β2AR, with and without the presence of the adrenergic inverse antagonist ICI118'551. B) mechanical response for HEK293G cell line overexpressing recombinant TS-SNAP-β2AR and HEK293 cells expressing endogenous (lower) level of receptor. C) Mechanical response in HEK293 cells expressing endogenous (lower) level of receptor, with and without application of the adrenergic inverse antagonist ICI118'551. Adapted from (Cullum et al., 2024).

**Figure 4: Production of cAMP in hCMEC/D3 cells assessed by a cAMP reporter assay in response to bacterial adhesion or β-adrenergic ligands.** WTNm, wild-type 2C43 bacteria; WTNm propra, wild-

type 2C43 bacteria in the presence of 100 nM propranolol; PilE- Nm, 2C43 DPilE bacteria devoid of type IV pili. iso, 10 mM isoproterenol; iso - propra, 10 mM isoproterenol and 100 nM propranolol; ***p < 0.001. Data are expressed as average values ± SEM. Adapted with permission from Coureil et al (Coureuil et al., 2010).

**Figure 5: Mechanical effects on β-AR-mediated cAMP response in adult cardiomyocytes.** A) Representative confocal images (out of 6 cells from 3 independent experiments) of CM under isotonic conditions (upper inset) and after 20 minutes of exposure to swelling medium (lower inset), with an overlay of cell edges under both conditions (main); B) representative curve (out of 3 independent experiments, 42 cells control and 47 cells swelling) showing kinetics of acceptor/donor ratio measured in CM stably expressing Epac1-camps under an epifluorescence microscope (normalized to baseline and 10 µM forskolin + 100 µM IBMX); Adapted from Sirbu et al. (Sirbu et al., 2024b).

**Figure 6: Effect of mechanical stimulation of β2-AR on adrenergic blockade.** A) Representative time-sequence (out of 3 independent experiments) of the relative increase of Nb80-eYFP fluorescence signal at the basolateral membrane, as isoproterenol is added after 60 min incubation with 100 µM propranolol; B) relative fluorescence increase measured upon Nb80-eYFP recruitment upon 1 µM and 10 µM isoproterenol stimulation in swollen vs non-swollen cells (mean ± SEM; n = 21 cells (control) and 23 cells (swelling) from 3 independent experiments; unpaired two-tailed t-test; p values: <0.0001 (1µM Iso), <0.0001 (10µM Iso) . Adapted from (Sirbu et al., 2024a).

**Figure 7: Mechanosensing effects on the signaling pathways of β2-AR and AT1R**. This figure depicts GPCR-mediated mechanotransduction, highlighting the downstream signaling pathways of β2-adrenergic receptor (β2-AR) and angiotensin II receptors (AT1R) in response to mechanical stimuli, as well as techniques for the measurement of these effects.

**Figure 8: OT deformations of cells expressing GPCR and cAMP biosensors.** A) Confocal image of a HEK293 cell expressing a fluorescently tagged membrane receptor, with a pair of tethers being pulled by a 2 μm bead using a Lumicks C-trap. B) A kymograph of a tether measured by a line-scan highlighted by the dashed line in panel A. The kymograph displays a fluorescent bleaching step followed by fluorescence recovery (FRAP). C) Profile of the fluorescent intensity recovery along the tether postbleaching that provides information on receptors diffusion rate on the tethers. D) Confocal section of a HEK293 cell expressing the cAMP biosensor PinkFlamindo. Here, upon application of mechanical pressure in a portion of the cell, by means of a trapped 2 μm bead, an increase in cAMP levels is recorded in the proximal region (A. Shetty and P. Annibale, unpublished data).

**Figures**

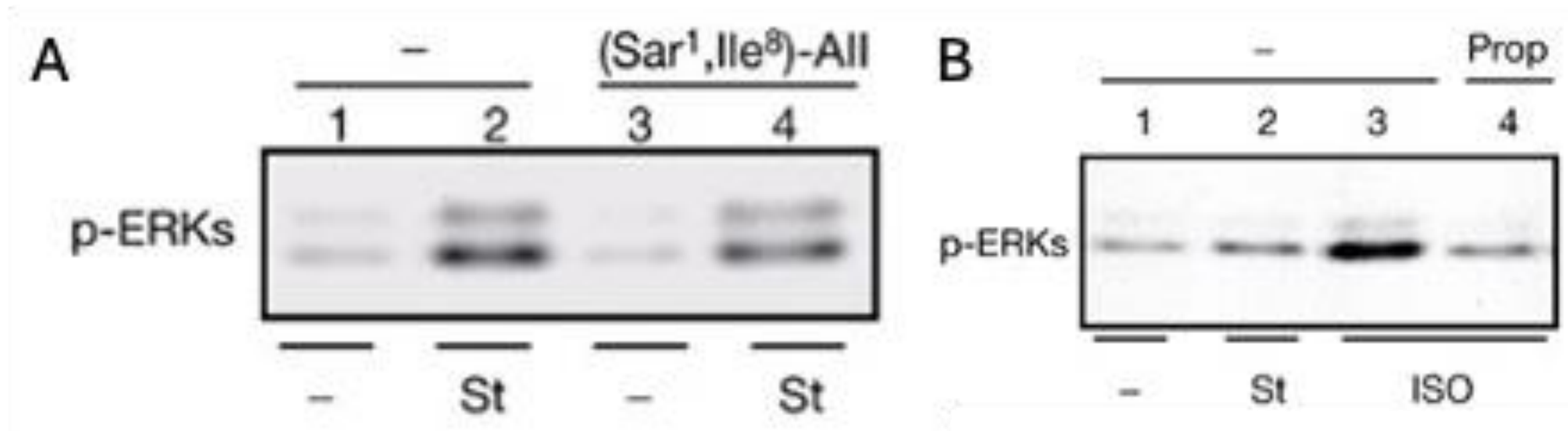

**Figure 1**

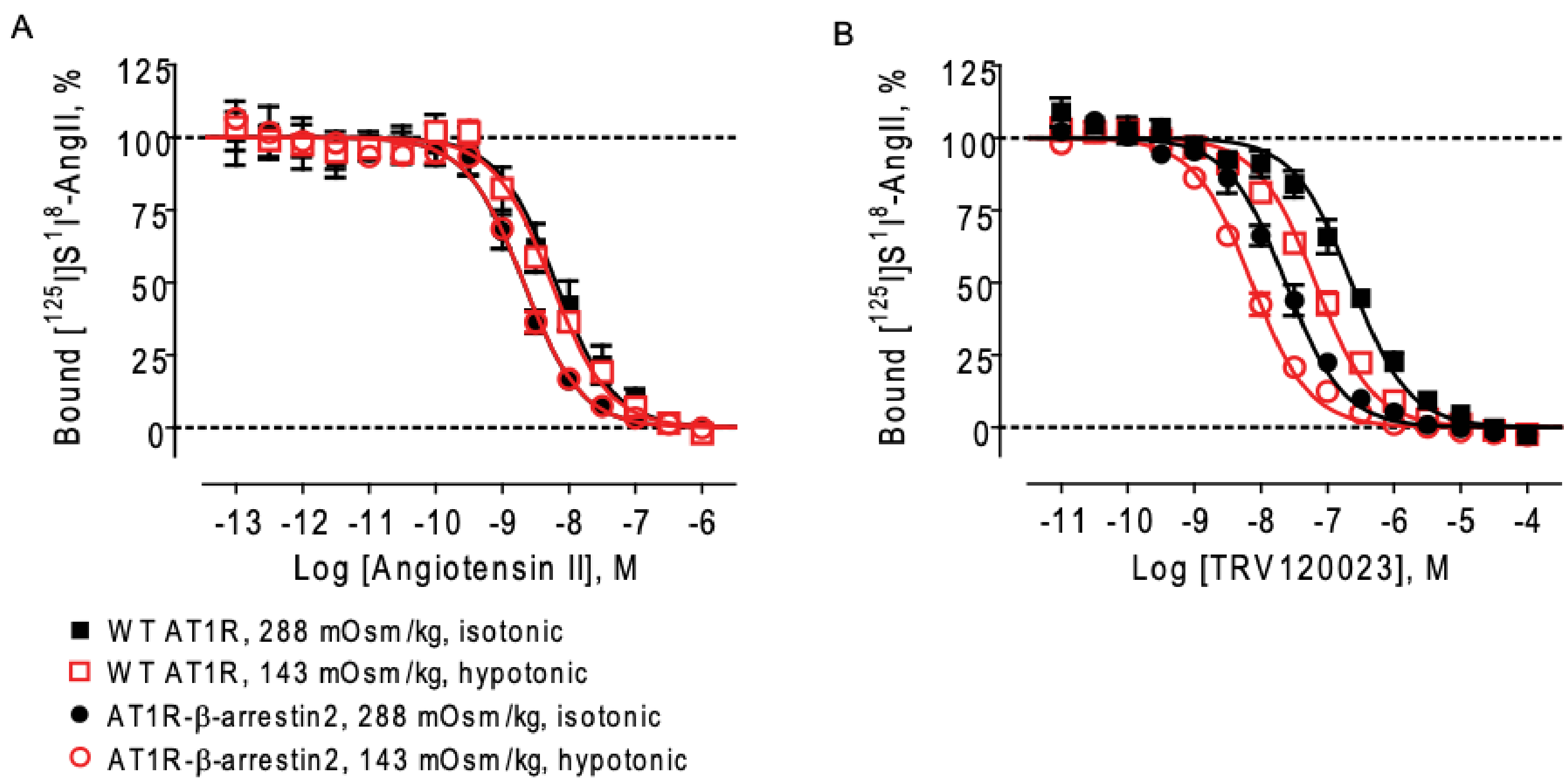

**Figure 2**

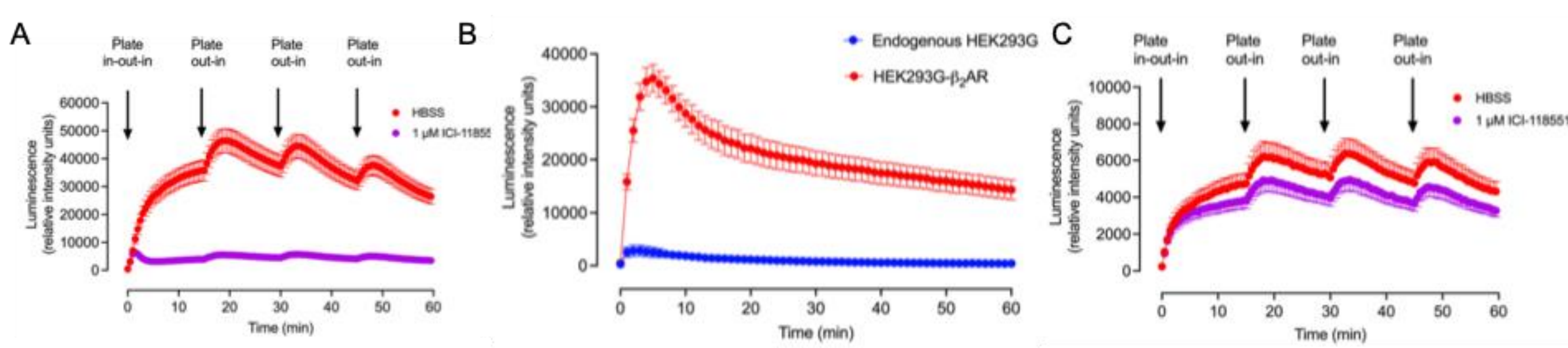

**Figure 3**

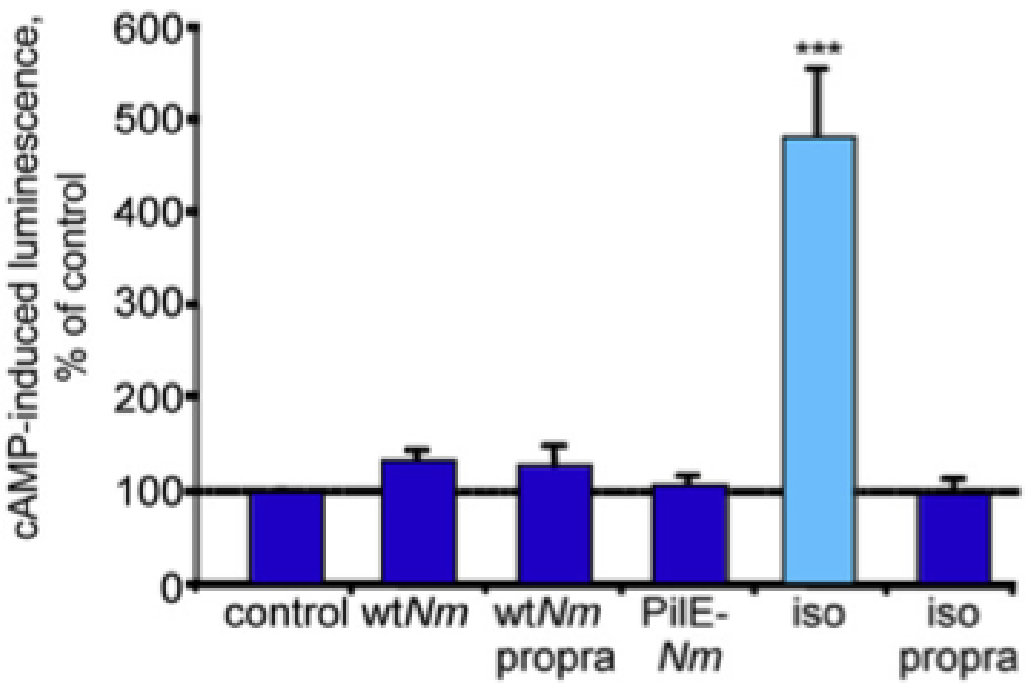

**Figure 4**

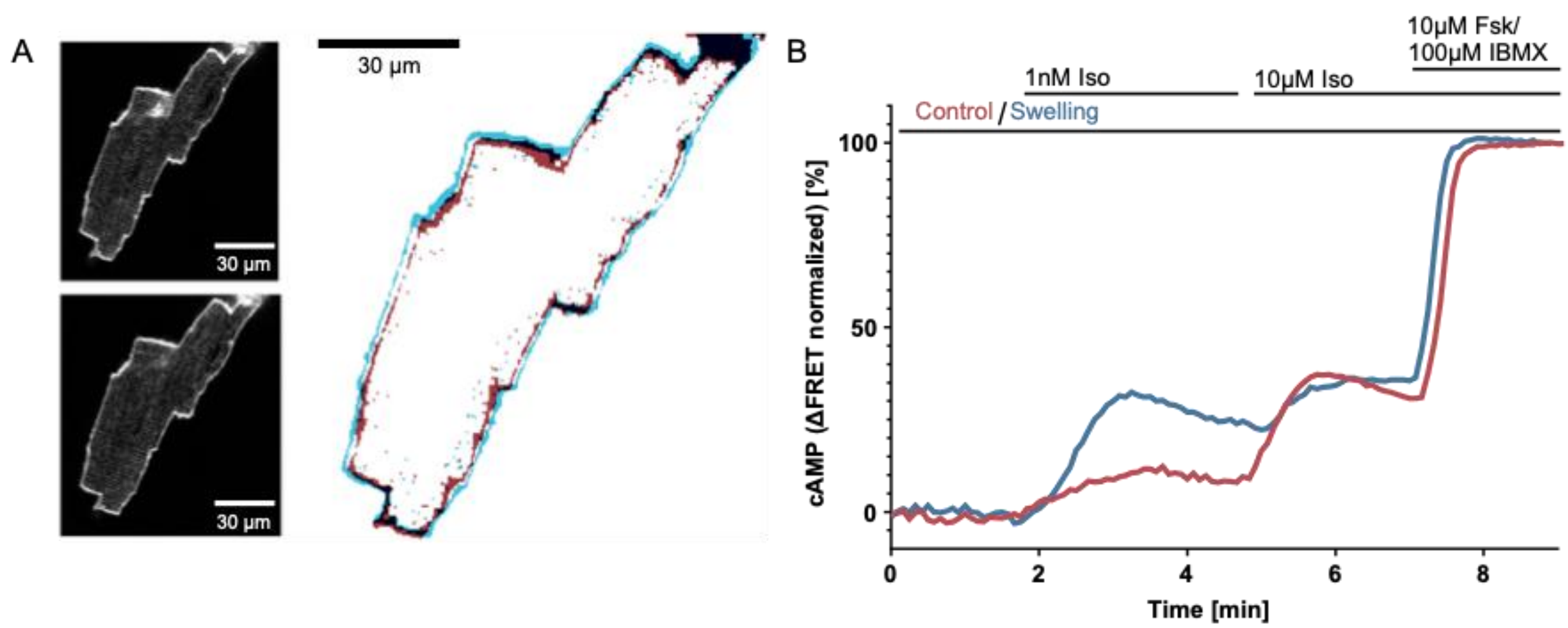

**Figure 5**

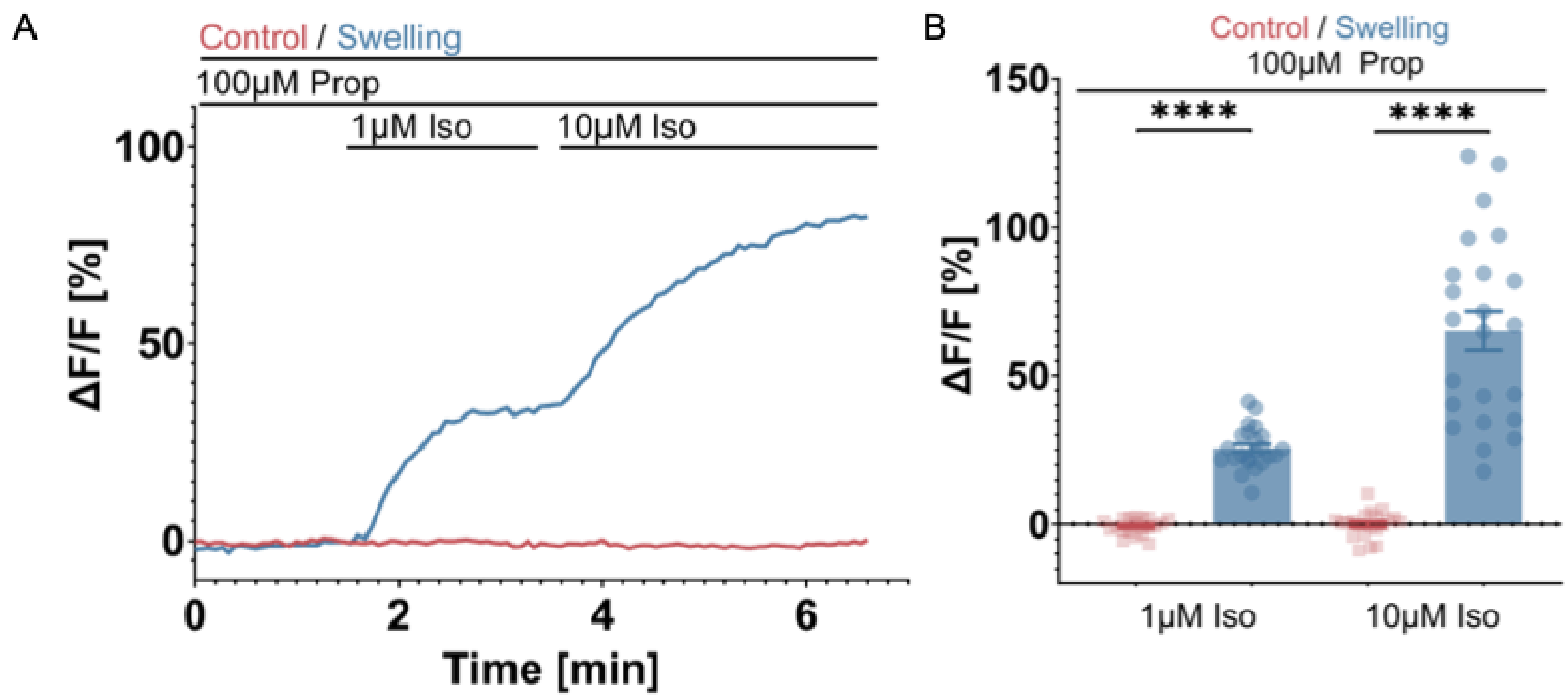

**Figure 6**

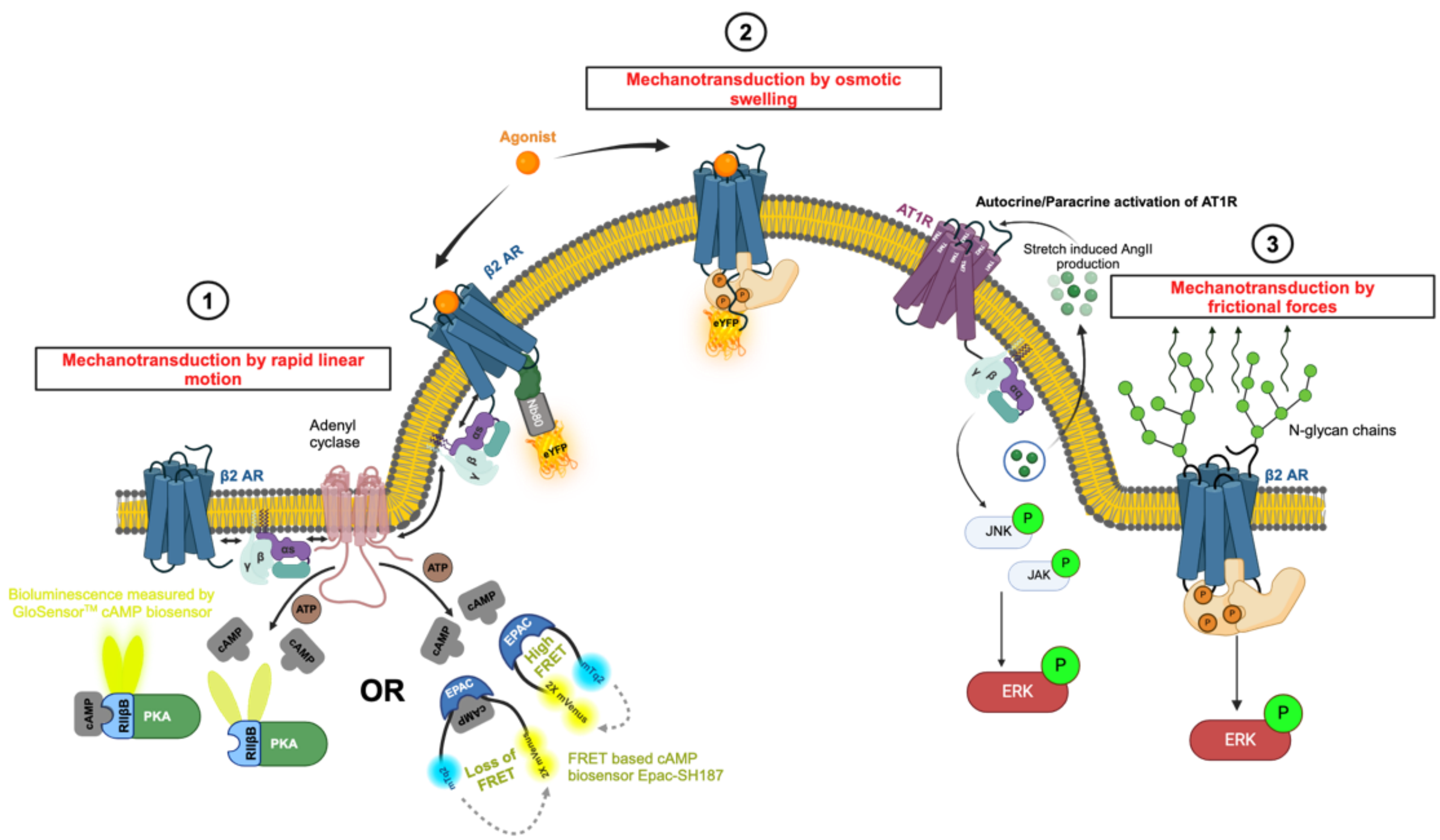

**Figure 7**

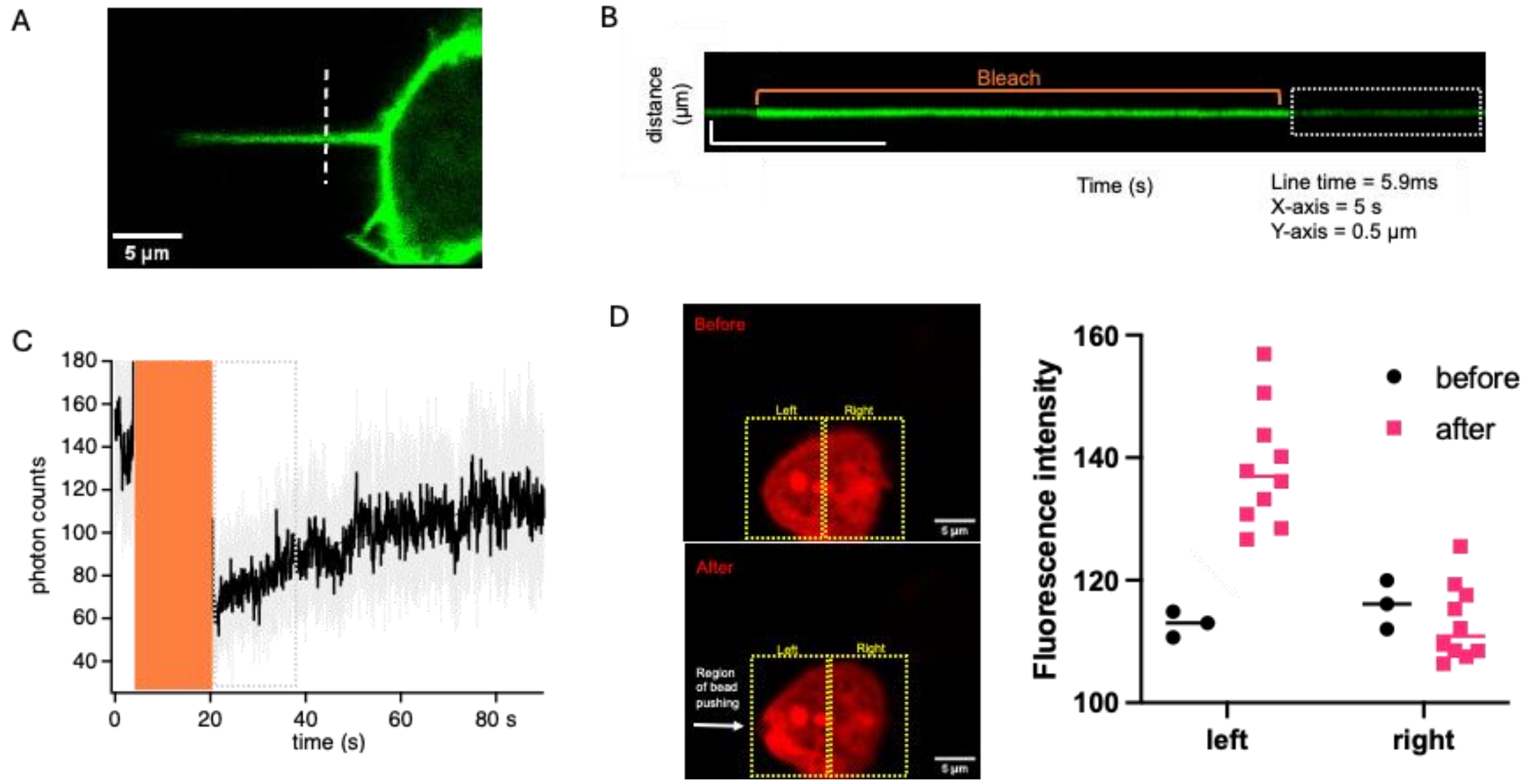
A
5 µm
B
Bleach
distance (µm)
Time (s)
Line time = 5.9ms
X-axis = 5 s
Y-axis = 0.5 µm
C
photon counts
180
160
140
120
100
80
60
40
0
20
40
60
80 s
time (s)
D
Before
Left
Right
After
Region of bead pushing
Fluorescence intensity
160
140
120
100
before
after
left
right

**Figure 8**